\documentstyle[pre,aps]{revtex}
\draft
\begin{document}
\title{
Flux flow resistivity and vortex viscosity of high-T$_{\rm c}$ films}
\author{J. I. Mart{\'\i}n$^\dagger$, M. V\'elez$^\dagger$,
F. Guinea$^\ddagger$ and J. L. Vicent$^\dagger$ \\}
\address{
	$\dagger$ Departamento de F{\'\i}sica de Materiales.
	Facultad de F{\'\i}sicas. Universidad Complutense. 28040 Madrid.
	Spain. \\
        $^\ddagger$ Instituto de Ciencia de Materiales. CSIC.
        Cantoblanco. 28049 Madrid. Spain.} 
\maketitle
\date{\today}
\begin{abstract}
The flux flow regime of high-T$_{\rm c}$ samples of 
different normal state resistivities is studied in the temperature range
where the sign of the Hall effect is reversed. The scaling of the vortex
viscosity with normal state resistivity is consistent with 
the Bardeen-Stephen theory. Estimates of the influence of  possible
mechanisms suggested for the sign reversal of the Hall effect 
are also given.
\end{abstract}

The dynamics of vortices in high-T$_{\rm c}$ materials 
show a rich behavior, not completely understood\cite{review}.
Part of the difficulties arise from the large variety of
defects which pin the vortices. Because of the efects induced
by pinning, it is a complicated task to separate intrinsic from
extrinsic effects.

One of the
most interesting topics in vortex dynamics is the anomalous sign
reversal of the flux-flow Hall effect\cite{Hall1,Hall2}. This anomaly cannot be
understood
within the usual Bardeen-Stephen model\cite{BS}. The alternative
analysis of vortex dynamics put forward by Nozi\`eres and Vinen\cite{NV}
cannot account for the phenomenon, either.
The study of the origin of this anomaly
has called the attention of many authors. One interesting point
of view is to explore the relationship between the longitudinal,
$\rho_{xx}$, and the transversal (Hall), $\rho_{xy}$, 
resistivities. This
have been done in the framework of scaling hypotheses
of the vortex dynamics\cite{scaling1,scaling2}.

In the present work, we study the flux flow regime, near T$_{\rm c}$,
of thin films of the 1:2:3 family, with different normal state resistivities. 
All of them display, in the same temperature range, a sign
reversal in the Hall resistivity. Our goal is to study the 
influence of the mechanism responsible for the sign reversal 
on the vortex viscosity. Samples with different normal state
resistivities are required to study the validity of the
standard Bardeen-Stephen theory\cite{BS} of vortex dissipation.
Note that, well below T$_{\rm c}$, it is generally accepted that 
the copper oxide superconductors are in the ultra clean limit, where
deviations from the Bardeen-Stephen expression are expected\cite{GP}.
Our results are consistent with those reported in\cite{fff},
where two samples with normal state resistivities $\sim 100 \mu \Omega$cm
were studied (see below).

Thin films of EuBa$_2$Cu$_3$O$_7$ have been grown on (100) 
SrTiO$_3$ substrates by dc magnetron sputtering, following 
standard procedures\cite{growth}. Samples are produced with the so
called c-axis texture (CuO$_2$ planes being parallel to the substrate).
Stoichiometric targets were used, the substrate target geometry was 
on-axis, the substrate temperature was, approximately, 800C during
deposition, the sputtering atmosphere was 85\% Ar and 15\%O up to a
total pressure of 300 mtorr. The annealing and cooling steps were
standard, as reported in the literature.

We chose three samples with different values of T$_{\rm c}$ and
normal state resistivities.  Normal state resistivities are taken
10K above the onset. Critical temperatures, with 
zero resistivity, varied between 80K and 90K. The samples were
patterned into regular bars (of width 500 $\mu$m and length 5 mm).
and the transverse (Hall) and longitudinal resistivities were
measured. Magnotransport effects were taken by a standard dc technique,
using a commercial 90 kOe magnet and temperature controller (Lake-Shore
DRC 91C). The Hall voltage was obtained from the antisymmetric 
part of the transverse voltage under magnetic field reversal.

Typical measurements are shown in figure (1). 
Results for the other two samples are given in
fig. (2). Our range of fields are comparable to
the ones used in\cite{fff}, and our results are consistent 
with the experiments reported there. Hence, we will use the 
measurements from\cite{fff} as a fourth case with different
normal state resistivity. Our results for three different samples 
are summarized in fig. (2). For comparison, results from
\cite{fff} are shown as circles (taken from sample II, with T$_{\rm c}$
= 88.5 K and $\rho_N \approx 100 \mu \Omega$ cm). 

The results are consistent with the Bardeen-Stephen theory of flux flow
viscosity. The values of $\eta \rho_N$ should scale with $H_{c_2}
\Phi_0 / c^2$, where $H_{c_2}$ is the upper critical field,
and $\Phi_0$ is the flux quantum.
As shown in fig.(3), the experimental data
is consistent with a linear dependence of $H_{c_2}$ on T - T$_c$,
with d$H_{c_2}$ / d T $\sim$ 2 Tesla / Kelvin.  Note that no adjustements
have been made on the available experimental data. The small variations
in d $H_{c_s}$ / d T suggest that the  superfluid
density changes little from sample to sample, as expected.

We now estimate the deviations from the Bardeen Stephen theory
arising from the recent proposal that vortices may be
charged\cite{Khomskii}, due to a shift in the chemical potential in the
superconducting state (see also\cite{vortex}). It has recently been
proposed that this effect can be measured at surfaces\cite{dipole}.

In order to analyze the contribution of the core charge to
the viscosity of a single vortex, we consider first an isolated 
pancake vortex, localized in a single CuO$_2$ plane. 
This point particle, as it moves under the influence of a voltage,
creates excitations in the medium, and dissipates energy.
Assuming that the leading cause of disspation is the creation of
electron-hole pairs, we can write the energy loss per unit time
as:

\begin{equation}
\frac{\partial E}{\partial t} = 
\eta | \vec{v} |^2 = \int d^D q V_q^2 {\rm Im} \chi
( \vec{q}, \vec{q} \vec{v} )
\label{eloss}
\end{equation}

where $\chi ( \vec{q} , \omega )$ is the polarizability 
due to electron-hole pairs of the
medium, and $V_q$ is the coupling between the core charge
and the electron-hole pairs. Neglecting for the moment the influence of
the superconducting gap, we know that, in a metal,
${\rm Im} \chi ( \vec{q} , \omega ) \propto { \omega } / { \epsilon_F^2 }$
where $\epsilon_F$ is the width of the conduction band of the metal.
From (\ref{eloss}), we can infer the value of the viscosity, $\eta$.
In general, for short range potentials, ( \ref{eloss} ) leads to
$\eta \sim {( \hbar V^2 )} / d({ \epsilon_F^2 a^2 )}$,
where $V$ is the potential induced by the ^^ ^^ impurity " on the
metal, and $a$ is its range, and $d$ is the separation
btween planes. This expression gives the
viscosity per unit length of the vortex.  Alternatively, we can replace
$V / \epsilon_F$ by $\delta$, the phaseshift induced by the
potential on the electrons at the Fermi level.
For the charged vortex considered here, $\delta$ should 
scale with the charge of the vortex, in dimensionless
units. Hence,
${V} / {\epsilon_F} \sim \delta \sim {Q} / {e}$.
The smallness of $Q$ justifies, a posteriori, the use
of second order perturbation theory in the present analysis
of the dissipation.
The range of the potential goes like the size of the core,
that is, the coherence length, $\xi$.
Finally, the vortex viscosity per unit length is:

\begin{equation}
\eta_Q \sim \frac{ \hbar Q^2}{e^2 \xi^2 d}
\label{etacharge}
\end{equation}

The standard theory of the stopping power of charges in
metals\cite{Lindhard,stop} gives a larger value for the
vortex viscosity per unit length, 
$\eta_Q \sim 0.1 Q^2 / d$ (in atomic units), for typical metallic
densities.  The main reason for this difference lies in the 
size of the potential due to the core charge,
which is taken to be of the order of the inverse 
Fermi-Thomas wavevector, $k_{FT}^{-1}$, in the second case.
A complete elucidation of this question requires a
detailed knowledge of the screening processes near the vortex
core\cite{dipole}.
The value of $\eta_Q$ is to be compared to the Bardeen-Stephen
contribution:

\begin{equation}
\eta_{BS} \sim \frac{\Phi_0 B_{c_2}}{\rho_N c^2}
\label{etabs}
\end{equation}

where $\Phi_O$ is the quantum unit of magnetic flux, $B_{c_2}$ 
is the upper critical field, $\rho_n$ is the normal state resistivity,
and $c$ is the velocity of light.

We can write $\Phi_0 B_{c_s} \sim B_{c_2}^2 \xi^2$ as
$\Delta F \xi^2$, where $\Delta F$ is the condensation
energy per coherence length to the cube, so that $\Phi_0 B_{c_2} \sim
\Delta^2 / ( \epsilon_F d )$, where $\Delta$ is the
superconducting gap. Using this last expression, and the value of 
$\eta_Q$ given in (\ref{etacharge}), we obtain:

\begin{equation}
\frac{\eta_Q}{\eta_{BS}} \sim \frac{\rho_N}{\left[ \frac{
\Delta^2 e^2 \xi^2}{\hbar Q^2 \epsilon_F c^2} \right]}
\label{ratio}
\end{equation}

We estimate the denominator in (\ref{ratio}) assuming that
$Q = 10^{-3} e , \epsilon_F = 1$eV, $\Delta = 0.05$eV and $\xi =
50$\AA. Then:
$\rho_Q = ({\Delta^2 e^2 \xi^2})/({\hbar Q^2 \epsilon_F c^2})
\approx 10^3 \mu \Omega {\rm cm}$.
If we use the standard expression for the stopping power of
a charge moving within a metal,  we obtain $\rho_Q \sim
10 \mu \Omega$ cm.
The relative importance of the vortex charge in the flux flow
dissipation can be inferred by comparing the value above to the normal
state resistivity of the material under consideration. 
The effect of the vortex charge will be important
if $\rho_Q < \rho_N$.
The two estimates given above can be considered as
an upper and a lower bounds, so that $10 \mu \Omega$ cm $<
\rho_Q < 10^3 \mu \Omega$ cm.
The samples studied
here have normal state resistivities within this range.

We now consider the possible sources of error in the derivation
of the estimate of $\rho_Q$ given earlier.
We assume that the response of the material is that of
a gapless metal. This is justified as far as
$\Delta \ll k_B T$, that is, near T$_{\rm c}$.
At lower temperatures, electron-hole pairs cannot be excited at
low energies, and the dissipation is reduced.
The other main approximation made in estimating $\rho_Q$ lies in 
the neglect of the temperature dependence of $Q, \Delta$ and $\xi$.
Note, however, that the product $\Delta \xi$ which enters in
$\rho_Q$ is independent of temperature. $Q$ goes to zero as
$T \rightarrow T_{\rm c}$, reducing the relative importance of
$\eta_Q$ near the transition temperature.  On general grounds,
$Q \sim \Delta^2 ( T ) \sim T_c - T$\cite{Khomskii}, and
$\rho_Q \sim ( T_c - T )^2$, which is not consistent with
the results shown in fig.(4).
Outside the critical region,
our estimate $Q \sim 10^{-3} e$ per plane\cite{Khomskii} is probably too
conservative\cite{dipole}. In any case, the value of $Q$ is the 
most uncertain parameter in $\rho_Q$.

In conclusion, we have analyzed the flux flow regime of high-T$_{\rm c}$
samples in the range where the sign reversal of the Hall effect is
observed. Samples with different normal state resistivities were used,
in order to verify the validity of the standard Bardeen Stephen theory
of flux flow dissipation.  Our results are consistent with this theory,
with d $H_{c_2}$ / d T $\approx$ 2 Tesla / Kelvin. Estimates of
the expected deviations associated to the charging of the vortices 
suggest that this effect should influence dissipation in samples
with normal state resistivities similar to those of the samples
studied here.  We find, however, no significant deviations from
the Bardeen-Stephen theory within our experimental errors.

We are grateful to prof. P. Echenique for helpful discussions.
This work has been supported through grant MAT94-0982, and grant
MAT96-0904 (CICyT, Spain).

\begin{figure}
\caption{Longitudinal resistivity, $\rho_{xx}$, Hall resistivity,
$\rho_{xy}$, and vortex viscosity, $\eta$, for a film with
T$_{\rm c}$ = 90K and $\rho_N = 370 \mu \Omega$ cm.}
\end{figure}

\begin{figure}
\label{measurements}
\caption{Longitudinal resistivities (upper curve), and
vortex viscosities (lower curves) for samples with 
normal resistivities $\rho_N = 76 \mu \Omega$ cm and 
T$_{\rm c} =$ 88K (a),
and $\rho_N = 800 \mu \Omega$ cm and T$_{\rm c}$ = 80.5K (b).}
\end{figure}

\begin{figure}
\label{eta}
\caption{Values of $\eta \rho_N$ as function of T$_{\rm c}$ - T for the three
samples described in the text. Circles are data taken from [9].}
\end{figure}

\end{document}